\documentclass[sigconf]{acmart}
\usepackage{times}  
\usepackage{helvet}  
\usepackage{courier}  
\usepackage{natbib}  
\usepackage{caption} 
\usepackage{algorithm}
\usepackage{algorithmic}
\usepackage{enumitem}
\usepackage{booktabs}
 \usepackage{array} 
\usepackage{amsmath}
\usepackage{multirow} 

\usepackage{newfloat}
\usepackage{listings}
\usepackage{xcolor}
\DeclareCaptionStyle{ruled}{labelfont=normalfont,labelsep=colon,strut=off} 
\floatstyle{ruled}
\newfloat{listing}{tb}{lst}{}
\floatname{listing}{Listing}
\title{An End-to-End Multi-objective Ensemble Ranking Framework for Video Recommendation}

\author{Tiantian He$^{*}$}
\affiliation{%
  \institution{Kuaishou Technology}
  \country{Beijing, China}
}
\email{hetiantian05@kuaishou.com}

\author{Minzhi Xie$^{*}$}
\affiliation{%
  \institution{Kuaishou Technology}
  \country{Beijing, China}
}
\email{xieminzhi@kuaishou.com}

\author{Runtong Li$^{*}$}
\affiliation{%
  \institution{Kuaishou Technology}
  \country{Beijing, China}
}
\email{lirt15@tsinghua.org.cn}

\author{Xiaoxiao Xu$^{*}$}
\affiliation{%
  \institution{Kuaishou Technology}
  \country{Beijing, China}
}
\email{xuxiaoxiao05@kuaishou.com}

\author{Jiaqi Yu$^{*}$}
\affiliation{%
  \institution{Kuaishou Technology}
  \country{Beijing, China}
}
\email{yujiaqi03@kuaishou.com}

\author{Zixiu Wang$^{*}$}
\affiliation{%
  \institution{Kuaishou Technology}
  \country{Beijing, China}
}
\email{wangzixiu@kuaishou.com}

\author{Lantao Hu$^{*}$}
\affiliation{%
  \institution{Kuaishou Technology}
  \country{Beijing, China}
}
\email{hulantao@kuaishou.com}

\author{Han Li$^{*}$}
\affiliation{%
  \institution{Kuaishou Technology}
  \country{Beijing, China}
}
\email{lihan08@kuaishou.com}

\author{Kun Gai$^{*}$}
\affiliation{%
  \institution{Unaffiliated}
  \country{Beijing, China}
}
\email{gai.kun@qq.com}

\usepackage{bibentry}

\begin{document}

\begin{abstract}

We propose a novel \textbf{E}nd-to-end \textbf{M}ulti-objective \textbf{E}nsemble \textbf{R}anking framework (EMER) for the multi-objective ensemble ranking module, which is the most critical component of the short video recommendation system. 
EMER enhances personalization by replacing manually-designed heuristic formulas with an end-to-end modeling paradigm.
EMER introduces a meticulously designed loss function to address the fundamental challenge of defining effective supervision for ensemble ranking, where no single ground-truth signal can fully capture user satisfaction.
Moreover, EMER introduces novel sample organization method and transformer-based network architecture to capture the comparative relationships among candidates, which are critical for effective ranking.
Additionally, we have proposed an offline-online consistent evaluation system to enhance the efficiency of offline model optimization, which is an established yet persistent challenge within the multi-objective ranking domain in industry.
Abundant empirical tests are conducted on a real industrial dataset, and the results well demonstrate the effectiveness of our proposed framework. In addition, our framework has been deployed in the primary scenarios of Kuaishou, a short video recommendation platform with hundreds of millions of daily active users, achieving a 1.39\% increase in overall App Stay Time and a 0.196\% increase in 7-day user Lifetime(LT7), which are substantial improvements.
\end{abstract}




\maketitle

\begingroup
\renewcommand\thefootnote{}\footnotetext{* These authors contributed equally to this work.}
\endgroup

\section{Introduction}

Short video sharing platforms attract a substantial number of users by providing valuable content.
When a user accesses a video platform, her satisfaction may manifest through diverse behaviors, which can be clustered into two groups. One group are user explicit interactions, such as liking, sharing, commenting, etc. The other group are user implicit feedback, such as watch time, long viewing, etc.
The ranking module is the most critical component of the recommendation system, which is ultimately responsible for determining which videos are presented to the user. 

\begin{figure}[h] 
    \centering
    \includegraphics[width=0.48\textwidth]{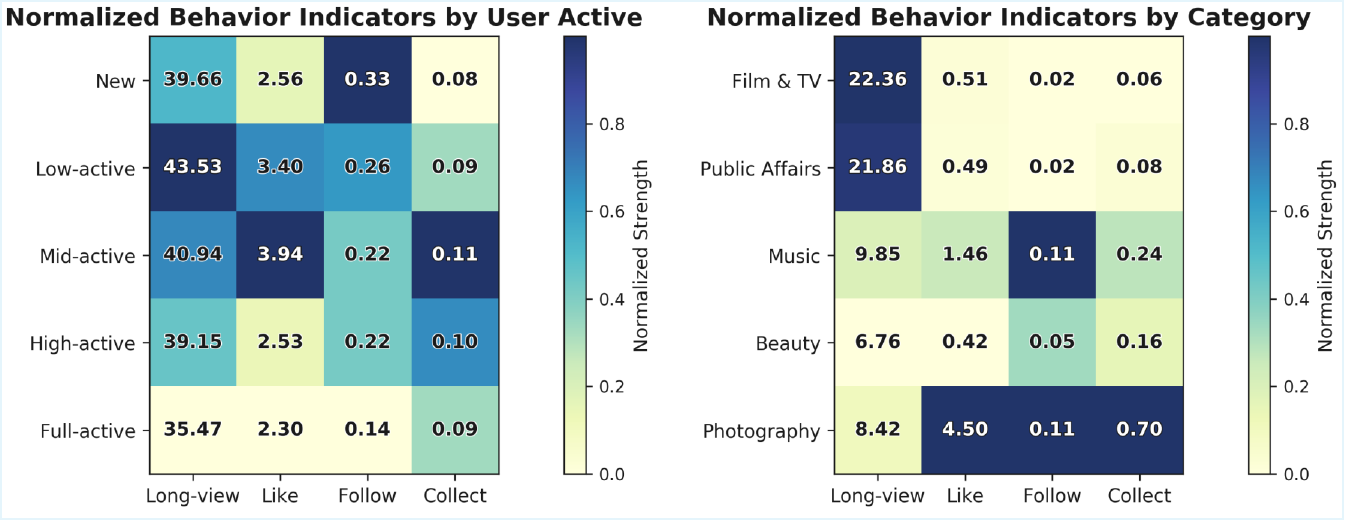} 
    \caption{Heatmaps of normalized user behavior indicators.
(left) User behaviors normalized by activity level, lower-activity users tend to express satisfaction through follows, with new users exhibiting approximately 135\% higher (0.33 vs 0.14) follow rates than fully active users.
(right) User behaviors normalized by content category, users demonstrate a higher likelihood of expressing satisfaction via ``longer viewing" 
for ``film/TV/entertainment" content, while preferring ``like, follow and collect" for music content. }
    \label{figure2}
    \vspace{-1em}
\end{figure}

The paradigm for ranking modules widely adopted in industry is a two-stage process. 
The first stage employs large-scale models to estimate user satisfaction in multiple dimensions. 
The second stage consolidates these multiple objectives into a single scalar value used for final ranking.
In industry, scalarization is predominantly achieved through manually designed heuristic formulas. Although these formulas offer advantages such as low implementation cost, flexible adjustment, and strong interpretability, they exhibit significant limitations in personalization and modeling complex non-linear interactions.

To enhance the personalization and nonlinear modeling capbility of multi-objective ensemble ranking modules, the industry has attempted to upgrade them to an end-to-end modeling paradigm. IntEL \cite{IntEL} explored a model-based approach; however, it retained reliance on a linear ranking formula, with the formula weights being learned by the model. 
A fundamental limitation of IntEL is the inherently constrained nonlinear capability of the linear formula.
UREM subsequently sought to augment IntEL by introducing an unsupervised loss function, aiming to maximize the model's perceptual capacity regarding input information. 
Nevertheless, existing research lacks a comprehensive discussion of the challenges related to practical deployment.

The core challenges in developing an end-to-end multi-objective ranking model are threefold:
1) Defining effective supervision signals: As illustrated in Figure \ref{figure2}, user satisfaction manifests through heterogeneous behavioral patterns that vary significantly across both user profiles and content taxonomies. That is to say, there is no single ground-truth signal can fully capture user satisfaction. In other words, unlike click-through rate prediction task, there is no unified ground-truth label for multi-objective ensemble ranking task.
2) Ensuring consistency between offline metrics and online performance: Objectives such as watch time and user interactions are susceptible to confounding factors. In our practical experience of optimizing user experience, we consistently observe a discrepancy: Offline model evaluations show improvements in AUC metrics for both watch time and interactions, yet online deployment yields divergent outcomes: increased watch time accompanied by a decline in interaction metrics.
3) Modeling comparative preferences among candidate items: Traditional recommendation system models estimate a user's satisfaction level with a given individual video. However, in real-world recommendation scenarios, the focus shifts to the ranking among multiple candidate videos, that is, to the comparative relationships between the videos within the candidate set. 

In this paper, we propose an end-to-end multi-objective ensemble ranking framework to address the above three core challenges.
To address the challenge of designing loss function, we introduce the concept of ``Relative Advantage Satisfaction", a personalized, comparative measure of user preference. We implement this metric via a pairwise ranking loss.
Furthermore, to improve ranking performance across multiple objectives, we incorporate a multi-dimensional AUC-based auxiliary loss. During training, an Advantage Evaluator dynamically adjusts the weights of this AUC loss in real-time, constraining the model towards ``self-evolution" in the direction of comprehensively superior AUC performance.
To address the issue of inconsistency between offline and online evaluations for watch time and user interactions, we propose a novel interaction metric: Interaction Probability per Unit Time (\textbf{IPUT}).
Through leveraging the backdoor criterion from causal inference, the \textbf{IPUT} metric eliminates the confounding factor of watch time inherent in interactions.
Furthermore, we designed a sophisticated sample organization scheme that aggregates all candidate videos to rank within a single request into one consolidated sample. A Transformer-based model architecture is specifically designed to process this aggregated set of candidate videos per request as its input, thereby enabling the model to capture the comparative relationships among the candidate videos.
We summarize the main contributions as follows:
\begin{itemize}[leftmargin=10pt]
\setlength{\itemsep}{-1pt}
    \item As far as we know, we are the first to clarify that the core challenges in model-based personalized multi-objective ranking are threefold: 1) the inherent difficulty in defining supervision signals, 2) establishing an offline-online consistent evaluation system, and 3) modeling comparative preferences among candidate items.
    \item To address the three core challenges in model-based personalized multi-objective ranking, we propose a comprehensive framework that includes the design of the loss function, the sample organization scheme, the model architecture, and the evaluation methodology.
    \item We verify the effectiveness of our proposed framework through extensive experiments conducted on a large scale industrial dataset, and successfully apply it in a real-world large scale recommender system bringing a considerable performance improvement.
\end{itemize}

\section{Related Work}
\subsection{Short Video Recommendation}
User-generated content platforms featuring short videos have increasingly become an integral part of daily life.  Recently, increasing research has focused on optimizing short video recommendation systems to enhance the accuracy and efficiency of recommending user-preferred content. Deep sequential models, especially GRU \cite{chung2014empirical} and Transformer architectures \cite{c:22}, have been popular for modeling user behaviors. They enhance interest modeling by capturing patterns in users' historical interactions \cite{hidasi2015session,zhou2018deep,zhou2019deep,pi2020search}.
Subsequently, research has shifted toward integrating multi-behavioral data to infer satisfaction more accurately. MIND \cite{li2019multi} retrieves multiple interest vector representations of users for retrieval, while ComiRec \cite{cen2020controllable} optimizes for diversity. Additionally, other research examines user interests from perspectives such as session analysis \cite{lv2019sdm} and multi-scale time \cite{tan2021dynamic}.
To jointly capture signals from diverse user feedback and optimize for multiple objectives,multi-task learning frameworks such as MMOE \cite{ma2018modeling} and PLE \cite{tang2020progressive} have been proposed.
An alternative approach to better understand user preferences is to collect explicit feedback through surveys \cite{christakopoulou2022reward}.
In recent developments, researchers have begun incorporating long-term interest \cite{pi2019practice,pi2020search,yan2024trinity,yi2023progressive} into these models to reflect the evolution of user interests over time. 
In this paper,  we evaluates ``relative user satisfaction" by examining various behaviors and their dynamic changes. By continuously refining this measure, we aim to better approximate users' true satisfaction levels and improve long-term personalization.

\subsection{Ranking Ensemble}
Modern recommendation systems often deploy multiple ranking models simultaneously to capture different aspects of user preferences. To combine the outputs into a unified result, ranking ensemble techniques have been widely explored. These techniques aim to leverage the complementary strengths of individual models to improve overall ranking performance.
Ranking ensemble methods are generally divided into unsupervised and supervised categories. Unsupervised methods follow predefined rules to merge rankings without requiring annotated data. Typical examples include Median Rank Aggregation \cite{fagin2003efficient}. In industrial systems, it is common practice to design weighted fusion formulas that integrate multiple ranking objectives into a unified score. These formulas typically have fixed structures, while the parameters such as objective weights are often manually set based on heuristics or prior knowledge. Some studies have explored search-based strategies like grid search and the Cross-Entropy Method to identify more effective configurations under offline evaluation metrics. Recently, growing efforts have been made to develop and optimize supervised rank ensemble methods.
Particularly, model-based techniques, such as the IntEL \cite{IntEL}, have demonstrated strong potential in learning effective rank aggregation strategies. Other approaches—such as those based on reinforcement learning \cite{zhang2022multi}, Differential Evolution \cite{balchanowski2022aggregation, balchanowski2022collaborative}, and genetic programming methods like Evolutionary Rank Aggregation \cite{oliveira2016evolutionary}—have also been investigated.
Collectively, these approaches highlight the increasing diversity of algorithmic strategies developed to enhance ranking ensemble performance in modern recommendation systems.


\section{Proposed Method}
\begin{figure*}[t] 
    \centering
    \includegraphics[width=0.93\textwidth]{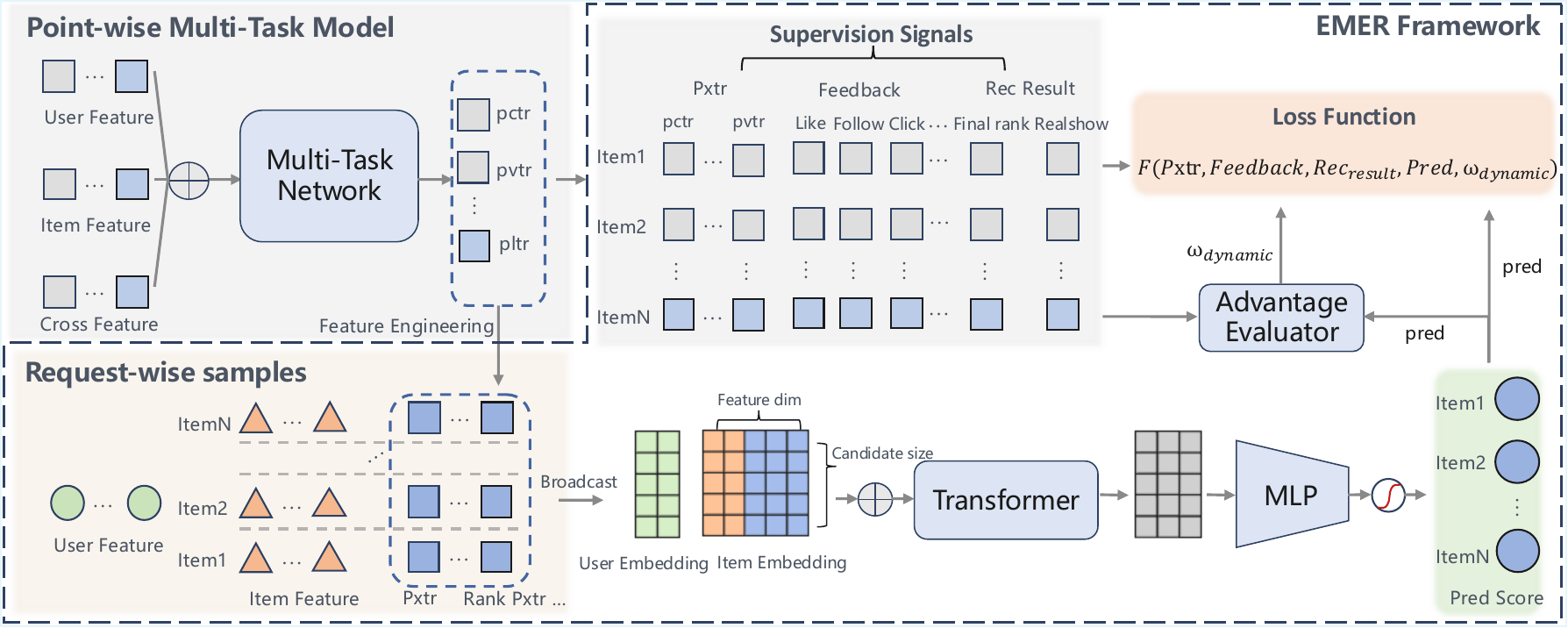} 
    \caption{An illustration of the EMER framework.}
    \label{figure3}
\vspace{-1.5em}
\end{figure*}

In this section, we provide a comprehensive overview of EMER. Firstly, we describe the details of EMER framework, including model architecture and loss function design, which is illustrated in Figure \ref{figure3}. Afterwards, we introduce an offline-online consistent evaluation system that enhances the efficiency of model optimization.

\subsection{EMER framework}

\subsubsection{Sample Construction \& Feature Encoding}

The essence of ranking lies in determining candidates' relative positions through mutual comparison. EMER highlights the comparative relationship information via the designated feature engineering method and the transformer-based model architecture, based on the introduction of a specific sample organization scheme. 

\textbf{Sample Organization:} 
When organizing the training data, we focus on two key goals: 1) Overcoming exposure bias inherent in learning solely from observed items. 2) Facilitating straightforward item comparison for the model. To achieve these, we meticulously organize training samples at the individual user request level, which means that all candidate items of a request, including exposed and unexposed items, are grouped together to form a consolidated training sample as shown in Figure ~\ref{figure3}.

\textbf{Feature Engineering:} To empower EMER's ability to model comparative relationships, we design features to capture each item's relative position within the candidate set. NormalizedRanks for different Pxtrs are introduced to provide explicit positional information, helping the model understand how the items stack up against the others in the given set, where 

$\mathrm{Normalized Ranks}=\frac{ \text{original rank}}{\text{total number of candidate items}} $

\subsubsection{Model Architecture}

EMER employs a transformer-based network to explicitly capture the complex relationships among candidates. This enables the model to weigh each item's influence on others, thereby better understanding their relative comparisons within the set.
In such way, the final predicted score reflects both intrinsic item quality and its relative standing within the candidate set.

\subsection{Learning Objectives \& Optimization Strategy}

Defining effective loss function is a significant challenge, as user satisfaction is pretty difficult to quantify uniformly. EMER introduces a label named ``Relative Advantage Satisfaction” based on the comparable post-exposure feedback within a recommendation request, and presents a solution from a multi-objective optimization perspective to concurrently enhance the model's ranking capability across multiple dimensions of user satisfaction. A self-evolving training scheme is introduced to ensure non-degraded ranking capability across all dimensions.

\subsubsection{Modeling Relative Advantage Satisfaction Labels}

Relative advantage satisfaction is defined based on posterior feedback (such as likes, follows, watch time et al.) observed after user receiving recommendations,  which often serve as the key evidences for evaluating user satisfaction. Given the inherently personalized and varied nature of user behavior, quantifying an absolute score from user feedback is exceptionally difficult. EMER addresses this challenge by defining satisfaction via relative comparisons, identifying the ``better" option based on user feedback.

A crucial principle is the hierarchy of ``Many Positives" over ``Single Positive" over ``No Positive": 1) ``Many Positives" refers to instances where a user interacts with a recommended item in multiple positive ways. For example, a user might not only $like$ an item (one Positive) but also $share$ it. 2) ``Single Positive"  occurs when a user provides one distinct type of positive feedback for a recommended item. This could be a $like$ or a $share$ on a video. 3) ``No Positive"  signify that the user provides no positive feedback on an item or even explicitly dislikes it.

For an individual user, relative satisfaction increases with the quantity of positive feedback received, and decreases as the number of negative feedback grows. Following the principle, pairwise logistic loss is employed to encourage pred $\hat{y}_{i}$ to be significantly higher than $\hat{y}_{j}$ for all positive pairs $D$:

\vspace{-1em}
\begin{eqnarray}
    L_{\text{posterior}} &=& -\sum_{(i,j) \in D} \log(P(x_{i} \triangleright x_{j})) \nonumber \\
    &=& -\sum_{(i,j) \in D} \log(\text{sigmoid}(\hat{y}_{i} - \hat{y}_{j}))
\end{eqnarray}





Although posterior feedback offers valuable insight into user satisfaction, it suffers from two intrinsic drawbacks: exposure bias (arising from models learning solely from shown items) and sparse positive signals (due to the infrequent nature of explicit user feedback). To mitigate these limitations, we introduce a complementary solution from a multi-objective optimization perspective.

\begin{figure}[h] 
    \centering
    \includegraphics[width=0.45\textwidth]{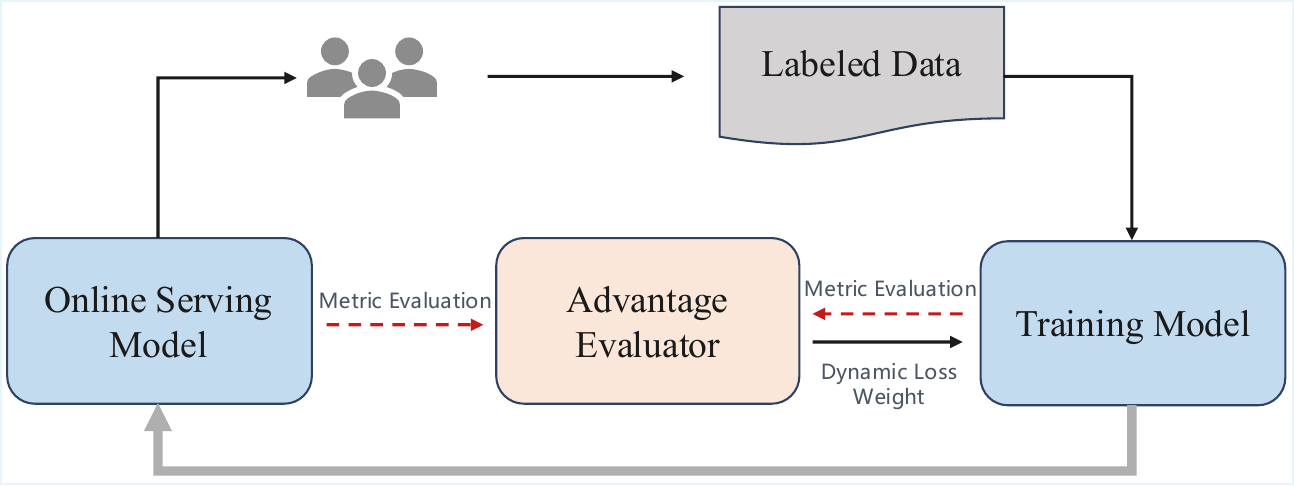} 
    \caption{End-to-end pipeline with online self-evolution.}
    \label{fig:wide1}
\vspace{-0.1em}
\end{figure}

\subsubsection{Enhancing multi-dimension ranking capability}

In large-scale recommendation systems, multi-task model (Figure~\ref{figure3}) provides rich prior signals ($Pxtrs$). The availability of these diverse signals for every candidate is crucial because it enables the model to effectively compare all candidates. Integrating these prior indicators into our user satisfaction definition framework allows for a more robust and forward-looking estimate of satisfaction.

However, effectively synthesizing and optimizing across such a conflicting set of objectives presents a significant challenge for most multi-objective ensemble ranking framework. Instead of combining all prior signals into one label, EMER treats each prior signal as a distinct supervision label, aiming to optimize ranking performance for each. This is predicated on the assumption that each signal reflects a unique facet of user satisfaction. Consequently, improving individual prior signal performance directly improves corresponding satisfaction facet. Therefore, systemic improvements across all prior signals collectively lead to heightened overall user satisfaction.

To achieve improvements across all prior signals, we need to find: 1)A proper loss function with direct alignment with ultimate ranking goal; 2)An effective way to deal with conflicts in multi-objective learning.

To align the learning objective with the ranking goal, EMER targets Area Under Curve (AUC), a standard metric for ranking quality in recommendation. However, direct optimization of AUC is intractable due to its non-differentiability and non-decomposable nature. To address this, we adopt a differentiable pairwise surrogate loss ~\cite{cortes2003auc, calders2007efficient}. This formulation approximates AUC maximization by encouraging the predicted score of a more preferred item to be higher than a less preferred one. The label is designed as:
\begin{equation}
 y_{pxtr_{\text{ij}}} = \left\{ \begin{array}{ll} 1 & \text{if }\, pxtr_{i} \triangleright pxtr_{j} \\ 0 & \text{otherwise} \end{array} \right. \
\end{equation}
where $pxtr$ represents one of $Pxtrs$. Positive sample is labeled 1 for item pair when $item_{i}$ has higher $pxtr$ than $item_{j}$ for each user.

The differentiable, pairwise surrogate loss for AUC encourages the probability $P_{i,j}$ that the predicted score $\hat{y}_{i}$ to be significantly higher than $\hat{y}_{j}$ for all positive samples.
\begin{equation}
    P_{i,j} = \text{sigmoid}(\hat{y}_{i} - \hat{y}_{j})
\end{equation}
\begin{equation}
    L_{pxtr} = -\sum_{(i,j) \in D} y_{pxtr_{\text{ij}}}\log(P_{i,j}) + (1-y_{pxtr_{\text{ij}}})\log(1-P_{i,j}) 
\end{equation}
where $L_{pxtr}$ is the loss for a single indicator $pxtr$. The total loss for all $Pxtrs$ is calculated as:
\begin{equation}
    L_{prior} = -\frac{1}{N}\sum_{pxtr}^{Pxtrs} w_{pxtr} \cdot L_{pxtr} 
\label{equ:aucloss}
\end{equation}
where $N$ denotes the number of $Pxtrs$, $w_{pxtr}$ is the weighting coefficients for different objectives (discussed in Self-Evolving Optimization Scheme).

Notably, prior-based satisfaction indicators offer two key advantages as supervision signals. First, they provide denser, continuous values—unlike sparse, binary feedback—which allows the model to learn more nuanced satisfaction levels. Second, they are unconstrained by exposure, eliminating the inherent exposure bias of observed feedback by providing a more comprehensive and unbiased signal for all items.




By combining prior and posterior satisfaction signals, we achieve a more complete and robust definition of user satisfaction. The final training loss is:
\vspace{-0.3em}
 \begin{equation}
    Loss = L_{posterior} + L_{prior}
\end{equation}


\subsubsection{Self-Evolving Optimization Scheme}

Approximating towards optimal solution for multi-objective is crucial, as different objectives often exhibit inherent conflicts. In EMER, we propose a ``Self-Evolving'' optimization scheme, employing an ``Advantage Evaluator'' to personally and dynamically compute the weights of the loss ($w_{pxtr}$ in Equation~\ref{equ:aucloss}), as shown in Figure~\ref{fig:wide1}.

\textbf{Self-Evolving:}
EMER introduces a novel concept of model self-evolving, which goes beyond static model updates to enable continuous and adaptive improvement. Unlike traditional approaches where models are retrained periodically on new data or with fixed optimization objectives, EMER is designed to ``learn to learn'' and dynamically adjust its internal mechanisms based on real-time performance and evolving user behaviors.

Consider a standard supervised learning scenario where a model $f(x; \Theta)$ aims to map an input $x$ to a output $\hat{y}$. The model's performance is quantified by a loss function $L(f(x; \Theta), y)$ over a dataset $D = \{(x_i, y_i)\}_{i=1}^{N}$ with label $y$. The optimization process seeks to minimize the loss by iteratively updating the parameters $\Theta$:

\vspace{-1em}
\begin{equation}
\Theta_{t+1} = \Theta_{t} - \eta \nabla_{\Theta} \mathcal{L}(\Theta_{t})
\end{equation}
where $\eta$ is the learning rate, $\mathcal{L}(\Theta)$ represents the  multi-objective loss function (equivalent to $L_{prior}$ in Equation ~\ref{equ:aucloss}) with $w_{pxtr}$ as weighting coefficient to balance the contributions of different objectives. 

Unlike methods relying on static weighting coefficients, Self-Evolving scheme employs completely personalized, dynamic and adaptive weighting coefficients. This is achieved by incorporating the performance of both present model $f(x; \Theta_{t})$ and previous model $f(x; \Theta_{t-1})$ into the current weight calculation.

\vspace{-1em}
\begin{equation}
w_{pxtr} = {\text{AE}}(f(x; \Theta_{t}); f(x; \Theta_{t-1}))
\end{equation}
where  $w_{pxtr}$ is strictly positive and $AE$ is the Advantage Evaluator defined in the paragraph below.
Notably, EMER does not require to save old model version to get $f(x; \Theta_{t-1})$. We exploit the fact that the current serving model reflects $\Theta_{t-1}$ due to online continual training, while the latest model with updated gradients represents $\Theta_{t}$.

\textbf{ Advantage Evaluator:}
 As mentioned before, Advantage evaluator doesn't measure absolute performance. It quantifies the \textit{relative advantage} of current model compared to historical baselines. 

\vspace{-1em}
\begin{equation}
\text{AE}(\Theta_{t}; \Theta_{t-1})= \frac{Metric(f(x_i; \Theta_{t-1}), y_i)}{Metric(f(x_i; \Theta_t), y_i)}
\end{equation}
where  $Metric(.)$ is the quantitative measure used to evaluate the ranking performance. Choosing one or multiple proper evaluation metrics is also crucial. Different tasks and business goals demand different ways of measurements.  Focusing on the  highly relevant items ranking performance, we try some different metrics (discussed in Ablation Studies).

Our self-evolving scheme provides a highly efficient and robust multi-objective optimization process. It intelligently eliminates the need for manual parameter tuning by dynamically and adaptively weighting each objective based on its performance. This fine-grained tuning operates at a per-request level, allowing the model to continuously adapt to personalized user preferences and shifting behaviors. The time-based self-comparison further enhances robustness by ensuring fair and accurate weight computation regardless of the inherent scale differences between objectives.






\subsection{Evaluation methodology}

The offline-online inconsistency problem is a tough challenge within the multi-objective ranking domain in industry. This issue arises when a model that performs exceptionally well on offline evaluation metrics fails to achieve similar performance gains when deployed in online environment.
This happened in our experiment: models consistently showed better AUC for both watch time and interactions in offline evaluations. But once deployed, we noticed watch time went up, while interactions actually dropped.


This gap highlights a fundamental issue we call the Decoupling Paradox which is deeply rooted in the confounding factor of watch time inherent in interactions.

Specifically, models are typically trained offline to optimize the cumulative interaction probability ($pxtr$) for individual items, aiming to maximize the likelihood of a user interacting with each separate item. However, in the online environment, the true goal is to maximize total cumulative interactions within a user's entire session. To achieve this, the model needs to increase the ``probability density" of interactions, striving to generate as many effective interactions per unit of time as possible.



 Consider a practical example: Assume that a user totally spends 20 seconds in a session and there are three candidates. Each is represented by (id, p\_watchtime, p\_interation): (1, 20s, 0.8), (2, 10s, 0.7), (3, 5s, 0.4). The pxtr-optimal sort for interaction ($0.8 > 0.7 > 0.4$) yields 0.8 expected interactions, while the probability density sort for interaction ($0.4/5s > 0.7/10s > 0.8/20s$) delivers at least 1.1 expected interactions. This example highlights that the offline pxtr is not a direct proxy for the "probability density" required online. This fundamental misalignment in objectives means that even if a model performs exceptionally well on offline pxtr, it cannot guarantee a proportional increase in online total interactions.

 To eliminate the confounding factor of watch time inherent in interactions, we reformulate the prior interaction $pxtr$ to novel Interaction Probability per Unit Time ($IPUT$) 

\vspace{-1em}
\begin{equation}
pxtr_{IPUT}=\frac{pxtr}{p\_watchtime}
\vspace{-0.3em}
\end{equation}

These new indicators shift the focus from merely an interaction likelihood to quantifying how efficiently that interaction occurs within a given time budget. This offers significantly more consistent performance between offline evaluations and online deployment.

\section{Experiment}

In this section, we compare the performance of EMER framework to previous ensemble methods. First, we introduce the basic settings of our ranking model. 
Then, we demonstrate the advantages of our method through both the performance of an online A/B test and the analysis of intermediate results.
Finally, we highlight the necessity of each module in EMER through an ablation analysis. 

\subsection{Experimental Setup}
Our model is employed in the ranking stage of two primary scenarios on the short video platform. 
The training dataset consists of recommendation logs generated while users engage with the short video application. Each sample represents a recommendation interaction, including a (user, candidate videos) pair, along with the corresponding feedback. 
The model is trained incrementally, processing over ten billion samples per day, with each sample containing approximately 500 candidate videos. 
The training is performed using the Adam optimizer with an initial learning rate of $ 5\times 10^{-6} $.
We evaluate the model's performance against two baseline methods:

\begin{itemize}
    \item Fusion formula (FF). The current online ensemble method of the recommendation platform. The ranking is determined by the score of each item, calculated using the formula $ score= f(\boldsymbol{\theta}_1, \mathtt{pvtr}) \times f(\boldsymbol{\theta}_2, \mathtt{pctr}) \times \cdots $ , where $ f $ is a simple smooth function and $ \{\boldsymbol{\theta}_1, \boldsymbol{\theta}_2, \cdots  \}$ represents the parameters of the function. Both the function and its parameters are manually configured according to the online performance.
    \item UREM~\cite{yu24}. An ensemble model that combines ranking similarity and reconstruction loss in an unsupervised manner to compute the final ranking score for each item.
\end{itemize}

\subsection{Overall Performance}
We evaluate the methods based on their performance in online A/B tests. A range of user satisfaction
metrics are encompassed as follows:
\begin{itemize}
    \item retention metric: 7-day Lifetime (LT7).
    \item implicit feedback metrics: APP stay time, watch time and video view. 
    \item explicit feedback metrics: actions such as like, follow, comment, forward, etc.
\end{itemize}

\begin{table}[h]
\caption{\textbf{A/B Performance Comparison between EMER and the Baseline Method.} The values in the table represent the improvement of EMER over the baseline method in the online A/B test.}
\label{tab:online-ab-result}
\small
\centering
\small 
\begin{tabular}{l|cc}
\toprule
{\textbf{Metrics}} & {\textbf{Scenario\#1}} & {\small\textbf{Scenario\#2}} \\
& {EMER vs FF} & {EMER vs FF}   \\
\midrule
LT7	&	+0.196\%	&	+0.133\%	\\
AppStayTime	&	+1.392\%	&	+1.199\%	\\
WatchTime	&	+1.542\%	&	+2.283\%	\\
VideoView	&	+1.044\%	&	+2.996\%	\\
Like	&	+0.435\%	&	-0.270\%	\\
Follow	&	+0.212\%	&	+0.381\%	\\
Comment	&	+0.695\%	&	+0.643\%	\\
Forward	&	+0.330\%	&	+1.038\%	\\

\bottomrule
\end{tabular}
\end{table}

As shown in Table~\ref{tab:online-ab-result}, EMER method significantly outperforms baseline approaches across most metrics. Notably, EMER achieves an increase of over $0.10\%$ in LT7 and more than $ 1.0\% $ in APP stay time in both scenarios. These are remarkable gains in industrial recommendation systems, demonstrating the effectiveness of EMER in video ranking tasks. The tendencies of LT7 are included in the supplement.

To systematically evaluate these ranking methods, we conduct an analysis of rank consistency between the rankings generated by ensemble methods and those derived from pxtrs. We consider the following various pxtrs that score videos from different perspectives.

\begin{itemize}
    \item pvtr: predicted watch time of a video.
    \item pctr/plvtr/pcpr: predicted probability of effectively/long/completely watching a video.
    \item pltr/pwtr/pcmtr/pftr: predicted probability of liking/following (the author of) /commenting/sharing a video.
    
\end{itemize}

To measure the consistency between two rankings, we choose the indicator $ \mathrm{GAUC}=\frac{\sum_i^{\#\text{user}} \#\text{impression}_{u_i} \cdot \mathrm{AUC}_{u_i}}{\sum_i \#\text{impression}_{u_i} } $, where $ u_i $ denotes the $i$-th user, and $ \#\text{impression}_{u_i} $ and $ \mathrm{AUC}_{u_i} $ represent the corresponding number of impressions and AUC respectively. GAUC is a common applied indicator in recommendation systems.
Table~\ref{tab:intermediate-result} indicates that our method has better consistency with most of the input pxtrs.

\begin{table}[h]
\caption{GAUC between Rankings Generated by Ensemble Methods and Pxtrs.}
\label{tab:intermediate-result}
\vspace{-0.5em}
\small
\centering
\begin{tabular}{l|ccc|ccc}
\toprule
\multirow{2}{*}{\textbf{\normalsize pxtr}} 
& \multicolumn{3}{c}{\textbf{Scenario\#1}} & \multicolumn{3}{c}{\textbf{Scenario\#2}} \\
& {\scriptsize EMER} & {\scriptsize UREM} & {\scriptsize FF} & {\scriptsize EMER} & {\scriptsize UREM} & {\scriptsize FF}  \\
\midrule
\textbf{pvtr}	&	\bf{0.727}	&  0.663  &	0.596	&	0.711 	&	\bf{0.718} 	&	0.601 	\\
\textbf{pctr}	&	\bf{0.689}	&  0.683  &	0.596	&	\bf{0.740} 	&	0.726	&	0.601 	\\
\textbf{plvtr}	&	\bf{0.704}	&  0.686  &	0.612	&	0.758 	&	\bf{0.761} 	&	0.645 	\\
\textbf{pcpr}	&	0.617	&  \bf{0.683}  &	0.579	&	\bf{0.758} 	&	0.701	&	0.648 	\\
\textbf{pltr}	&	\bf{0.706}	&  0.611 &	0.556	&	0.648 	&	\bf{0.655}	&	0.575 	\\
\textbf{pwtr}	&	\bf{0.704}	&  0.593  &	0.561	&	\bf{0.683} 	&	0.629	&	0.567 	\\
\textbf{pcmtr}	&	\bf{0.706}	&  0.616   &	0.567	&	\bf{0.706} 	&	0.667	&	0.578 	\\
\textbf{pftr}	&	\bf{0.743}	&  0.624  &	0.580	&	\bf{0.736} 	&	0.687 	&	0.584 	\\

\bottomrule
\end{tabular}
\vspace{-1.5em}
\end{table}

\subsection{Further Analysis}

In this section, we further analyze the necessity of each module in EMER through Ablation Studies. Additionally, we briefly explain how our chosen metrics influence video recommendations in the Metric Study. 

\label{subsec:further_analysis}

\subsubsection{Ablation Studies}
\label{4.3.1}

To validate our proposed EMER framework, we conducted comprehensive ablation studies. By systematically removing or replacing key components, we quantify their individual contributions to both offline GAUC and online A/B test performance. Our findings, summarized in Tables~\ref{tab:offline_ablation} and~\ref{tab:online_ablation}, lead to the following conclusions.


\begin{table}[h] 
\vspace{-0.5em}
\scriptsize 
\centering
\captionsetup{aboveskip=5pt, belowskip=0pt} 
\captionsetup{width=\linewidth} 
\caption{Offline Ablation Studies in term of GAUC.} 
\vspace{-0.5em}
\label{tab:offline_ablation}
\setlength{\tabcolsep}{3pt}
\renewcommand{\arraystretch}{1.2}
\begin{tabular}{l|ccccccc}
\toprule
\textbf{Metric} & \textbf{\textit{Fusion}} & \textbf{\textit{EMER-}} & \textbf{\textit{EMER-}} & \textbf{\textit{EMER-}} & \textbf{\textit{EMER-}} & \textbf{\textit{EMER-}} & \textbf{\textit{EMER}} \\
& \textbf{\textit{Formula}} & \textbf{\textit{NoComp}} & \textbf{\textit{NoPost}} & \textbf{\textit{NoPrior}} & \textbf{\textit{NoEvolve}} & \textbf{\textit{NoIPUT}} & \\
\midrule
\textbf{pctr} & 0.596 & 0.668 & 0.653 & 0.642 & 0.688 & 0.683 & \textbf{0.689} \\
\textbf{pvtr} & 0.596 & 0.688 & 0.684 & 0.659 & \textbf{0.734} & 0.722 & 0.727 \\
\textbf{plvtr} & 0.612 & 0.685 & 0.679 & 0.661 & 0.701 & 0.700 & \textbf{0.704} \\
\textbf{pcpr} & 0.579 & 0.602 & 0.601 & 0.600 & 0.551 & 0.607 & \textbf{0.617} \\
\textbf{pltr} & 0.556 & 0.689 & 0.690 & 0.693 & 0.680 & 0.678 & \textbf{0.706} \\
\textbf{pwtr} & 0.561 & 0.697 & 0.681 & 0.691 & 0.684 & 0.680 & \textbf{0.704} \\
\textbf{pcmtr} & 0.567 & 0.681 & 0.686 & 0.692 & 0.619 & 0.686 & \textbf{0.706} \\
\textbf{pftr} & 0.580 & 0.732 & 0.723 & 0.721 & 0.631 & 0.709 & \textbf{0.743} \\
\bottomrule
\end{tabular}
\vspace{-0.5em} 
\end{table}


\begin{table}[h] 
\caption{Online A/B Test Performance over Fusion Formula.}
\scriptsize 
\centering
\captionsetup{aboveskip=2pt, belowskip=0pt} 
\captionsetup{width=\linewidth} 
\setlength{\tabcolsep}{2pt} 
\renewcommand{\arraystretch}{1.2} 
\begin{tabular}{l|cccccc} 
\toprule
\textbf{Metric} & \textbf{\textit{EMER-}} & \textbf{\textit{EMER-}} & \textbf{\textit{EMER-}} & \textbf{\textit{EMER-}} & \textbf{\textit{EMER-}} & \textbf{\textit{EMER}} \\
& \textbf{\textit{NoComp}} & \textbf{\textit{NoPost}} & \textbf{\textit{NoPrior}} & \textbf{\textit{NoEvolve}} & \textbf{\textit{NoIPUT}} & \\
\midrule
\textbf{AppStayTime} & +1.094\% & +1.045\% & +0.939\% & +1.233\% & +1.242\% & \textbf{+1.392\%} \\
\textbf{WatchTime} & +1.309\% & +1.253\% & +1.013\% & \textbf{+1.689\%} & +1.409\% & +1.542\% \\
\textbf{VideoView} & +0.953\% & +0.960\% & +0.802\% & -1.303\% & +1.001\% & \textbf{+1.044\%} \\
\textbf{Like} & -1.740\% & -0.539\% & -1.688\% & -1.494\% & -1.649\% & \textbf{+0.435\%} \\
\textbf{Follow} & -0.482\% & -1.525\% & +0.045\% & -1.393\% & -7.948\% & \textbf{+0.212\%} \\
\textbf{Comment} & -0.720\% & -1.178\% & -1.162\% & -7.414\% & -6.664\% & \textbf{+0.695\%} \\
\textbf{Forward} & -1.235\% & -1.448\% & -1.542\% & -8.088\% & -10.826\% & \textbf{+0.330\%} \\
\bottomrule
\end{tabular}
\label{tab:online_ablation}
\end{table}

\textbf{Ablation of Modeling Comparative Relationships:}
To validate the effectiveness of explicitly modeling candidates' comparative relationships, we created a variant called EMER-NoComp. This model removes the request-level sample grouping, the transformer-based network and $Normalized Ranks$ as input. EMER-NoComp scores each item in isolation rather than determining their relative positions. The results show a measurable decline in both offline metrics (Table~\ref{tab:offline_ablation}) and online metrics (Table~\ref{tab:online_ablation}). This confirms that modeling comparative relationships within the candidate set is crucial for enhancing item representation and ranking quality.

\textbf{Ablation of Components in Supervision Signals:}
EMER's supervision signals have two parts: posterior-based relative satisfaction and prior-based multi-dimensional objectives. We created two variants: EMER-NoPost (removes posterior-based signals) and EMER-NoPrior (removes prior-based signals). Both variants show a performance decline, confirming the individual effectiveness of each component. The decline is more significant with EMER-NoPrior, which validates the crucial role of prior signals due to their density, lack of exposure bias, and capacity to provide multi-dimensional insights.

\textbf{Ablation of Self-Evolving Optimization Scheme:}
To investigate the core contribution of the self-evolving scheme, we designed EMER-NoEvolve, a variant that trains using static weights for different objectives. As shown in Table~\ref{tab:offline_ablation} and Table~\ref{tab:online_ablation}, EMER-NoEvolve exhibits a significant performance imbalance across various metrics. For example, online WatchTime increases, but at the cost of significant drops in other metrics (e.g., VideoView -2.347\%, Forward -8.418\%, and Comment -8.109\%). This confirms that removing the self-evolving scheme greatly diminishes the model's ability to resolve multi-objective conflicts. Even with substantial manually weights tuning, EMER consistently outperformed all manually tuned configurations. A deeper analysis (Figure~\ref{fig:loss} in the Appendix) revealed that EMER's multi-objective loss distribution is more uniform and centralized with lower value than EMER-NoEvolve's. This clearly validates that the self-evolving scheme is crucial for guiding the model's learning direction and ensuring its robustness and stability over time.

\textbf{Ablation of Interaction Probability per Unit Time:}
As shown in Table~\ref{tab:mubd_ablation}, EMER-NoIPUT suffers from a classic decoupling paradox: significant offline improvements in the AUC of $pltr$ and $pwtr$ fail to translate into positive online A/B test results. In contrast, positive offline gains of $pxtr_{IPUT}$ directly align with positive online results. This ablation demonstrates that IPUT module is vital for aligning offline training with real-world online user satisfaction.

\begin{table}[h!] 
\small 
\centering
\captionsetup{aboveskip=5pt, belowskip=0pt} 
\captionsetup{width=\linewidth}
\caption{Offline-Online Consistent Evaluation Study.}
\vspace{-0.5em}
\label{tab:mubd_ablation}
\setlength{\tabcolsep}{5pt} 
\renewcommand{\arraystretch}{1.2} 

\begin{tabular}{lccc} 
\toprule
\textbf{Metric} & \textbf{\textit{Fusion Formula}} & \textbf{\textit{EMER-NoIPUT}} & \textbf{\textit{EMER}} \\
\midrule
\multicolumn{4}{l}{\textbf{\textit{Offline GAUC Metric}}} \\ 
\textbf{pltr} & 0.556 & 0.678 & \textbf{0.706} \\
\textbf{pwtr} & 0.561 & 0.680 & \textbf{0.704} \\
\textbf{pcmtr} & 0.567 & 0.686 & \textbf{0.706} \\
\textbf{pftr} & 0.580 & 0.709 & \textbf{0.743} \\
\textbf{pltr\textsubscript{IPUT}} & 0.555 & 0.516 & \textbf{0.567} \\
\textbf{pwtr\textsubscript{IPUT}} & 0.556 & 0.505 & \textbf{0.587} \\
\textbf{pcmtr\textsubscript{IPUT}} & 0.598 & 0.521 & \textbf{0.612} \\
\textbf{pftr\textsubscript{IPUT}} & 0.557 & 0.512 & \textbf{0.574} \\
\midrule 
\multicolumn{4}{l}{\textbf{\textit{Online A/B Test over FF}}} \\ 
\textbf{Like} & - & -1.649\% & \textbf{+0.435\%} \\
\textbf{Follow} & - & -7.948\% & \textbf{+0.212\%} \\
\textbf{Comment} & - & -6.664\% & \textbf{+0.695\%} \\
\textbf{Forward} & - & -10.826\% & \textbf{+0.330\%} \\
\bottomrule
\end{tabular}
\vspace{-0.5em}
\end{table}


\begin{table}[h] 
\small 
\centering
\captionsetup{aboveskip=1pt, belowskip=0pt} 
\captionsetup{width=\linewidth} 
\caption{Offline Metric Study in term of GAUC.} 
\label{tab:offline_metric} 
\setlength{\tabcolsep}{5pt} 
\renewcommand{\arraystretch}{1.2} 

\begin{tabular}{lcccccc} 
\toprule
\textbf{Metric} & \textbf{$HitRate@K$} & \textbf{$MEAN@K$} & \textbf{\textit{DCG@K}(ours)} \\
\midrule
\textbf{pctr}   & 0.654 & 0.674 & \textbf{0.689} \\
\textbf{pvtr}   & 0.699 & 0.711 & \textbf{0.727} \\
\textbf{plvtr}  & 0.681 & 0.683 & \textbf{0.704} \\
\textbf{pcpr}   & 0.593 & 0.584 & \textbf{0.617} \\
\textbf{pltr}   & 0.673 & 0.676 & \textbf{0.706} \\
\textbf{pwtr}   & 0.689 & 0.681 & \textbf{0.704} \\
\textbf{pcmtr}  & 0.670 & 0.688 & \textbf{0.706} \\
\textbf{pftr}   & 0.690 & 0.721 & \textbf{0.743} \\
\bottomrule
\end{tabular}
\vspace{-1em}
\end{table}

\subsubsection{Metric Study}

Within the Self-Evolving Optimization Scheme, the Advantage Evaluator critically assesses the current model and historical baselines. The choice of its evaluation metric is paramount. We investigated three strategies (Equation~\ref{equ:hr}-~\ref{equ:dcg}): $HitRate@K$, $MEAN@K$, and $DCG@K$ shown in the Appendix, using $pxtr$ as a positive signal. As Table~\ref{tab:offline_metric} shows, \textbf{$DCG@K$} consistently outperforms $HitRate@K$ and $MEAN@K$ across nearly all offline GAUC metrics. This indicates $DCG@K$, by emphasizing top-ranked item relevance and position, leads to more accurate predictions and better aligns with complex user engagement. Therefore, we adopted the $DCG@K$ as the default evaluation metric for our dynamic weighting.

\section{Conclusion}
In this paper, we present EMER, a novel end-to-end multi-objective ensemble ranking framework designed to enhance personalization in large-scale recommendation systems. EMER addresses the challenge of defining supervision signals for the ensemble ranking model by employing a meticulously designed loss function. Moreover, EMER introduces novel sample organization method and transformer-based network architecture to capture the comparative relationships among candidates, which are critical for effective ranking. Our framework has been successfully deployed in the main feed of Kuaishou, a short-video recommendation platform that serves hundreds of millions of daily active users, achieving substantial improvements in user engagement and overall system performance. These results demonstrate the practical utility and scalability of EMER in real-world industrial environments.

\bibliographystyle{ACM-Reference-Format}
\balance
\bibliography{main}

\appendix

\section{Appendix}

\begin{itemize}
    \item $HitRate@K$: The intersection of the top-K items from the pxtr ranking list and the final ranking list.
            \begin{equation}
            HitRate@K_{pxtr} = \frac{List_{pxtr, K} \cap List_{final, K}}{K}
            \label{equ:hr}
            \end{equation}
    \item $MEAN@K$: The mean $pxtr$ score of the top-K ranking($r$) items from the final ranking list.
           \begin{equation}
            MEAN@K_{pxtr} = \frac{\sum_{r=1}^{K} pxtr_{r}}{K}
           \label{equ:mean}
           \end{equation}
    \item $DCG@K$: The cumulative gain with a ranking positional ($r$) discount from the top-K final ranking list.
           \begin{equation}
            DCG@K_{pxtr} = \sum_{r=1}^{K} \frac{pxtr_{r}}{\log_2(r + 1)}
            \label{equ:dcg}
            \end{equation}
\end{itemize}

\begin{figure}[H] 
    \centering
    \includegraphics[width=1.0\columnwidth]{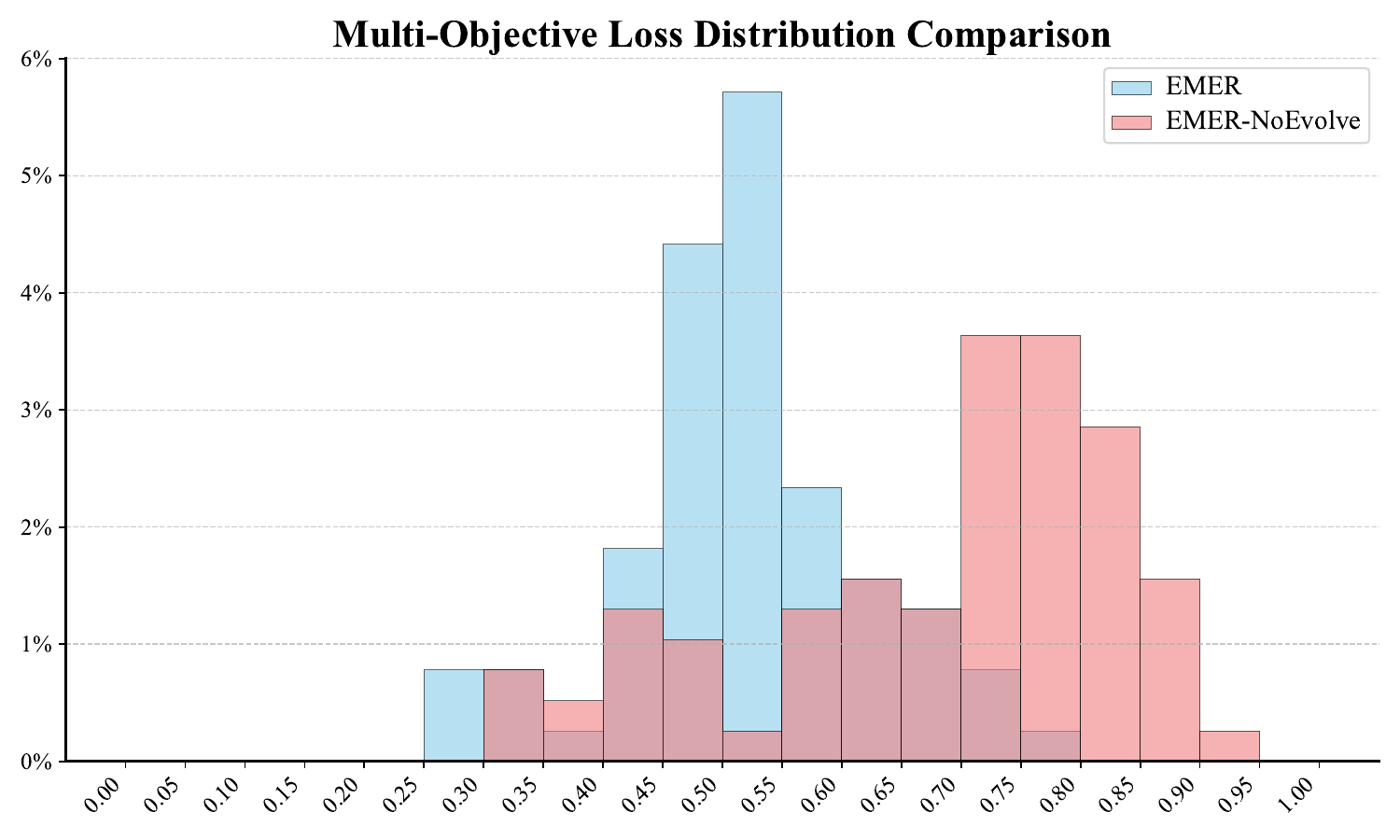} 
    \caption{Multi-Objective Loss Distribution Comparison. EMER achieves a lower, more uniform multi-objective loss distribution, while EMER without MSE leads to a skewed distribution with high-loss values.}
    \label{fig:loss}
\end{figure}

\begin{figure}[H] 
    \centering
    \includegraphics[width=1.0\columnwidth]{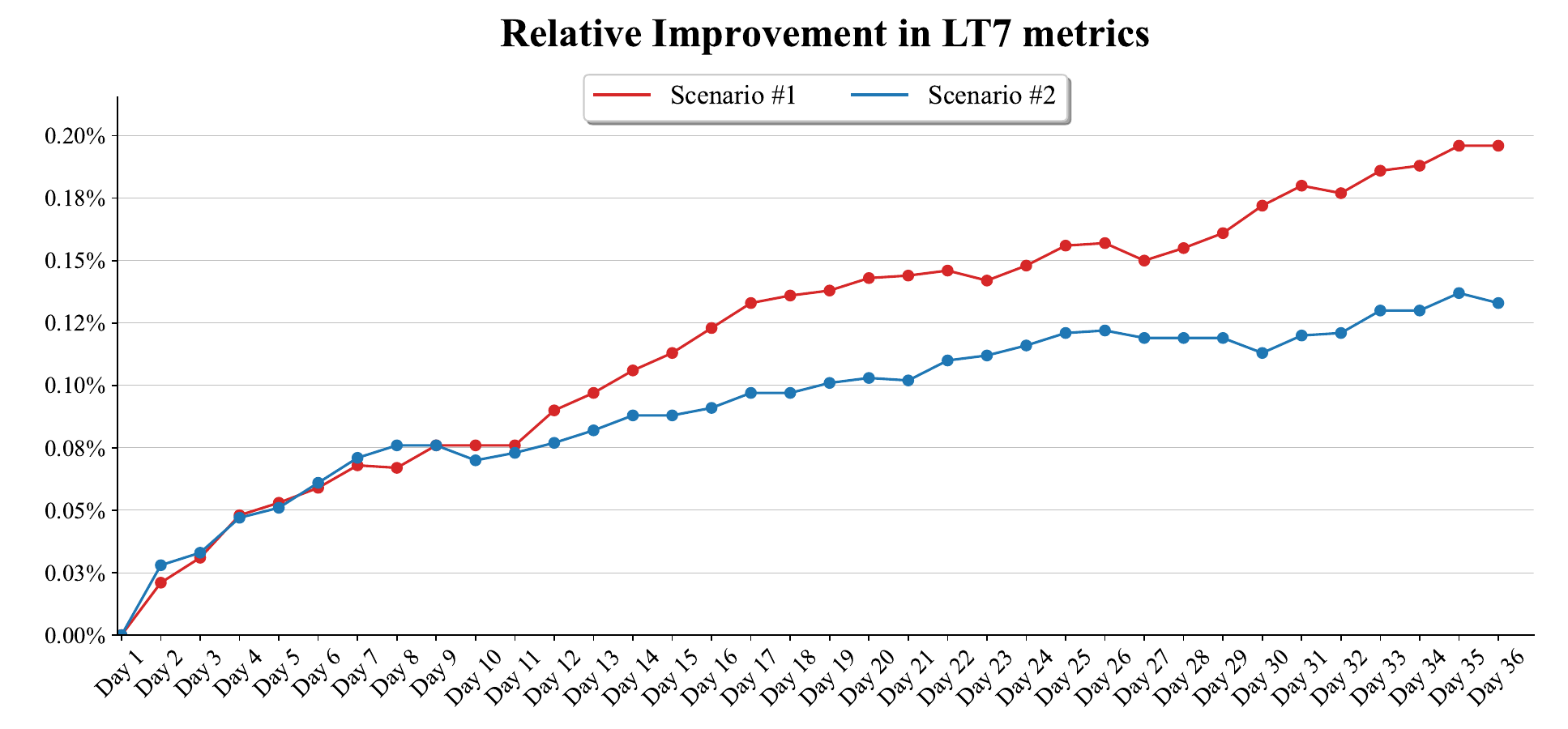} 
    \caption{Relative Improvement in LT7 metrics over Fusion Formula.}
    \label{fig:lt7}
\end{figure}

\end{document}